\documentclass[conference]{IEEEtran}
\IEEEoverridecommandlockouts

\usepackage{amsmath, amsfonts, amssymb, bm, amsthm}
\usepackage[utf8]{inputenc}
\usepackage{enumitem} 
\usepackage[ruled,vlined]{algorithm2e} 

\makeatletter
\renewcommand\@makefnmark{\hbox{\@textsuperscript{\normalfont\color{black}\@thefnmark}}}
\makeatother

\usepackage{stfloats}
\usepackage[bottom]{footmisc}
\usepackage{color, xcolor, soul, framed}
\usepackage{cancel}
\usepackage[numbers,sort&compress]{natbib}
\usepackage[font=normalfont]{caption} 
\usepackage{diagbox}
\usepackage{pifont}

\usepackage{tikz}
\usetikzlibrary{arrows.meta, positioning, calc, shapes.geometric}
\usetikzlibrary{decorations.pathreplacing}

\theoremstyle{remark}

\usepackage{subcaption}
\usepackage{booktabs}
\usepackage{mathtools}
\usepackage{graphicx}
\usepackage{hyperref}
\usepackage{varwidth}
\usepackage{amsmath}
\usepackage{multirow}
\usepackage{array}
\usepackage{lipsum}
\usepackage{orcidlink}

\definecolor{turquoise}{rgb}{.0,.3,1.0}

\usepackage[first=0,last=9]{lcg}

\newcommand{\CC}{\cellcolor{Gray!80!white}}

\newcommand{\CG}{\cellcolor{green!3!white}}
\usepackage{color, colortbl}
\definecolor{Gray}{gray}{0.93}

\hypersetup{
    colorlinks=true,    citecolor=blue,    linkcolor=red,    filecolor=magenta,  urlcolor=blue,    pdftitle={Overleaf Example},
    }

\begin{document}
\title{
\LARGE{\bf{AERO-LQG: Aerial-Enabled Robust Optimization for LQG-Based Quadrotor Flight Controller}}}
\author{Daniel~Engelsman and Itzik~Klein 
\thanks{The authors are with the Charney School of Marine Sciences, Hatter Department of Marine Technologies, University of Haifa, Haifa, Israel. 
\noindent E-mails: \{dengelsm@campus, kitzik@univ\}.haifa.ac.il}}

\maketitle
\begin{abstract}
Quadrotors are indispensable in civilian, industrial, and military domains, undertaking complex, high-precision tasks once reserved for specialized systems. Across all contexts, energy efficiency remains a critical constraint: quadrotors must reconcile the high power demands of agility with the minimal consumption required for extended endurance.
Meeting this trade-off calls for mode-specific optimization frameworks that adapt to diverse mission profiles. At their core lie optimal control policies defining error functions whose minimization yields robust, mission-tailored performance. While solutions are straightforward for fixed weight matrices, selecting those weights is a far greater challenge—lacking analytical guidance and thus relying on exhaustive or stochastic search. This interdependence can be framed as a bi-level optimization problem, with the outer loop determining weights a priori.
This work introduces an aerial-enabled robust optimization for LQG tuning (AERO-LQG), a framework employing evolutionary strategy to fine-tune LQG weighting parameters. Applied to the linearized hovering mode of quadrotor flight, AERO-LQG achieves performance gains of several tens of percent, underscoring its potential for enabling high-performance, energy-efficient quadrotor control.
The project is available @
\texttt{\url{http://github.com/ANSFL/AERO-LQG}}. %
\end{abstract}

\section{Introduction}
Unmanned aerial vehicles (UAVs), particularly quadrotors, have rapidly evolved from research prototypes into practical tools across a broad range of applications. Their versatility and agility enable complex missions—such as surveillance \cite{leahy2016persistent}, inspection \cite{liang2023towards}, delivery \cite{li2023drone}, and search and rescue \cite{lyu2023unmanned}—making them valuable assets in both civilian and military domains \cite{lewis2011drones}.
From a dynamical systems perspective, quadrotor flight modes span a wide stability spectrum—from near-equilibrium hovering to highly nonlinear, aggressive maneuvers. As a result, no single control strategy can achieve optimal performance across all regimes. Transitions between control modes are therefore common, especially as energy constraints impose competing demands: controllers must either deliver agility for high-intensity missions \cite{Hanover2024auto}, or conversely, limit maneuverability to conserve energy and extend flight time \cite{engelsman2025c, bauersfeld2022range}. 
\\
Monolithic control lacks such flexibility; adaptive scheduling with task-dependent adjustments is needed \cite{liew2017recent, elmeseiry2021detailed, telli2023comprehensive}. Frequent mode switching, however, complicates state estimation, which remains fragile to the quadrotor’s nonlinear and underactuated dynamics—despite advances in hardware \cite{el2021indoor}, inertial sensing \cite{engelsman2023information}, and AI-driven perception \cite{cohen2024inertial}. To mitigate these issues, linear quadratic Gaussian (LQG) control remains a staple in aerial robotics, valued for its theoretical rigor and optimality under specific assumptions \cite{valavanis2014handbook}.
\\
Many stabilized flight modes can be approximated as linear systems, allowing precise LQG control, provided operating conditions stay near the linearization point and modeling uncertainties are managed \cite{chrif2014aircraft}. Quadrotors are no exception—their hovering mode, in particular, can be effectively linearized to facilitate both control and state estimation.
However, while the quadratic form of the objective cost function is inherently convex for fixed weights, tuning these values constitutes an additional outer optimization task—highly non-convex due to the coupled estimator–controller dynamics \cite{saviolo2023learning}. This raises a key question:
\begin{quote}
Could bi-level optimization redefine the limits of LQG tuning for quadrotor stability ?
\end{quote}
\begin{figure}[t]
\centering
\begin{tikzpicture}[
  >=Latex,
  node distance=16mm and 24mm,
  block/.style = {draw, rounded corners=3pt, thick, align=center, inner sep=4pt, minimum width=14mm, minimum height=10mm},
  arrow/.style={-{Stealth[length=6pt, width=6pt]}, rounded corners=1.5mm, line width=1.25pt},
  labelnode/.style={font=\footnotesize, inner sep=1pt},
  every node/.style={font=\footnotesize}
]

\tikzset{jump/.style={preaction={draw=white, -, line width=8pt}, postaction={draw}, -{Stealth[length=6pt, width=6pt]}, line width=1.25pt}}

\node[block] (kf)  {Kalman\\filter};
\node[inner sep=0pt, right=15mm of kf, align=center] (quad) {\includegraphics[width=22mm]{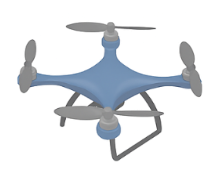}};
\node[block, above=3mm of quad] (ric) {Riccati\\equations};
\node[block, above=8mm of ric] (evo) {Evolutionary\\optimizer};
\node[block, below=14mm of quad] (perf) {Performance\\metrics};
\node[block, right=15mm of quad] (lqr) {LQR\\controller};


\draw[arrow] (ric.east) node[above, xshift=13mm] {LQR gain} -| (lqr.north);
\draw[arrow] (ric.west) node[above, xshift=-11mm] {KF gain} -| (kf.north);
\draw[arrow] (kf.south) -- ++(0,-12mm) node[right, yshift=4.5mm, xshift=1mm] {state estimates} -| (lqr.south);
\draw[arrow] (perf.west) node[above, xshift=-14mm] {responsiveness} -- ++(-34mm,0) |- (evo.west);
\draw[arrow] (perf.east) node[above, xshift=14mm] {tracking errors} -- ++(+34mm,0) |- (evo.east);
\draw[arrow] (evo.south) -- node[xshift=1.4mm] {LQG~~~weights}(ric.north);

\draw[arrow] (quad.west) --  node[xshift=1mm] {\shortstack{measur-\vspace{1.5mm}\\ ~~~ements}}(kf.east);
\draw[arrow] (lqr) --  node[above, xshift=1mm] {command}(quad.east);
\draw[jump] (quad.south) -- node[right, yshift=3mm, xshift=1mm] {action}(perf.north);
\end{tikzpicture}
\caption{AERO-LQG applies bi-level optimization: the outer loop uses an evolutionary strategy to optimize the weighting matrices, while the inner loop computes the LQG gains.}
\label{fig:intro}
\end{figure}
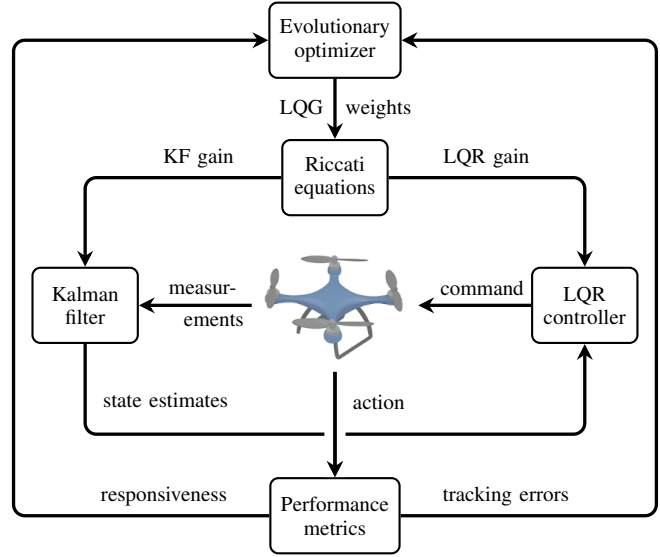
%
To answer this question, we develop an aerial-enabled robust optimization for LQG tuning (AERO-LQG), tailored to modern quadrotors. As illustrated in Fig.~\ref{fig:intro}, an outer evolutionary loop refines the weighting matrices, while an inner loop computes the LQG gains, together forming a nested architecture. Evaluated across diverse metrics, AERO-LQG delivers unprecedented gains—reducing tracking error by over 55\% and extending flight endurance by nearly 8\%.
\\ 
The paper unfolds as follows: Section~\ref{sec:theory} presents the problem setup and technical foundations; Section~\ref{sec:prop} outlines the proposed approach; Section~\ref{sec:results} discusses the findings; and Section~\ref{sec:conc} concludes the study.

\section{Problem Setup} \label{sec:theory}
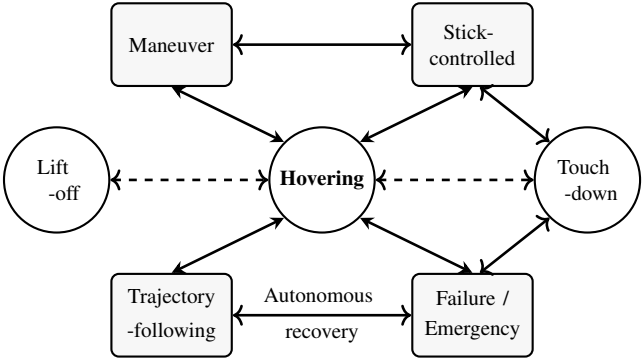
\begin{figure}[h]
\centering
\begin{tikzpicture}[
  box/.style={draw, thick, minimum width=16mm, minimum height=11mm, fill=gray!6, rounded corners=1mm, inner sep=1.5pt},
  circ/.style={draw, circle, thick, fill=gray!0, minimum size=14mm, inner sep=2pt},
  arrow-1/.style={-{Stealth[length=5pt, width=5pt]}, line width=1.pt},
  arrow-2/.style={<->, >={Stealth[length=5pt, width=5pt]}, line width=1.pt}, 
  labelnode/.style={font=\footnotesize, inner sep=1pt},
  every node/.style={font=\footnotesize}
]

\node[circ] (takeoff) {\shortstack{Lift \vspace{.5mm}\\ ~ -off}};
\node[circ, right=21mm of takeoff] (hold) {\textbf{Hovering}};
\node[circ, right=21mm of hold] (land) {\shortstack{Touch \vspace{.5mm}\\ \ -down}};
\node[box, shift={(-20mm,18mm)}] at (hold) (stabilize){Maneuver};
\node[box, shift={(+20mm,18mm)}] at (hold) (manual){\shortstack{Stick- \vspace{.5mm}\\ controlled}};
\node[box, shift={(+20mm,-18mm)}] at (hold) (emergency) {\shortstack{\ Failure / \vspace{1mm}\\ Emergency}};
\node[box, shift={(-20mm,-18mm)}] at (hold) (auto) {\shortstack{Trajectory\vspace{1mm}\\ -following}};

\draw[arrow-1, dashed, <->] (takeoff) -- (hold);
\draw[arrow-1, <->] (stabilize.east) -- (manual.west);
\draw[arrow-1, dashed, <->] (hold) -- (land);
\draw[arrow-2, <->] (stabilize.south) -- (hold.north west);
\draw[arrow-2, <->] (manual.south) -- (hold.north east);
\draw[arrow-1, <->] ([xshift=1mm]manual.south) -- (land.north west);
\draw[arrow-2, <->] (hold.south west) -- (auto.north);
\draw[arrow-2, <->] (hold.south east) -- (emergency.north);
\draw[arrow-1, <->] ([xshift=1mm]emergency.north) -- (land.south west);
\draw[arrow-1, <->] (emergency.west) -- node {\shortstack{Autonomous \vspace{2.5mm}\\ recovery}} (auto.east);
\end{tikzpicture}
\caption{State transition diagram with hovering as the central hub connecting diverse nonlinear flight modes (boxes).}
\label{fig:FMS}
\end{figure}

Quadrotor behavior is modeled as a continuous-time dynamical system, defined by the following differential equation
\begin{align}
\dot{\boldsymbol{x}}(t) = \boldsymbol{f}(\boldsymbol{x}(t), \boldsymbol{u}(t)) \ , \label{eq:sys}
\end{align}
where \( \boldsymbol{x}(t) \in \mathbb{R}^n \) is the state vector, \( \boldsymbol{u}(t) \in \mathbb{R}^k \) is the control input vector, and \( \boldsymbol{f}: \mathbb{R}^n \times \mathbb{R}^k \to \mathbb{R}^n \) is a smooth function that defines its dynamics. The solution to this system describes the trajectory of the state \( \boldsymbol{x}(t) \), starting from an initial condition \( \boldsymbol{x}(0) \) and evolving under the influence of the input \( \boldsymbol{u}(t) \). 
\\
Most aerial flight modes such as take-off, uncoordinated turns, maneuvers, and landing, require the system states to change dramatically, i.e. $\dot{\boldsymbol{x}}(t) \neq \mathbf{0}$, thereby causing $\boldsymbol{f}$ to evolve nonlinearly in time. That said, certain operating points, if controlled properly, enable maintaining the system at an equilibrium, denoted by subscript \( e \), with their states demonstrating negligible change \cite{arrowsmith1990introduction}, such that
\begin{align}
\dot{\boldsymbol{x}}(t) = \boldsymbol{f}(\boldsymbol{x}_e, \boldsymbol{u}_e) = \boldsymbol{0} \ . \label{eq:sys_eq}
\end{align} 
The equilibrium behavior is assessed using local linearity, approximating the nonlinear dynamics \( \boldsymbol{f} \) with a linear model near the equilibrium point \( (\boldsymbol{x}_e, \boldsymbol{u}_e) \). General fixed-wing configurations permit linear approximation of several trimmed (stable) flight modes—such as straight-and-level flight, coordinated turns, and steady climbs or descents—thanks to the gliding capability of their large wing area. Rotorcraft, by contrast, depend entirely on rapid, coordinated motor corrections, yielding predominantly nonlinear dynamics \cite{saviolo2023learning, peksa2024review}. Consequently, the only flight mode that consistently maintains equilibrium under small perturbations is hovering (center), as depicted in the finite state machine in Fig.~\ref{fig:FMS}.

\subsection{Equilibrium Point}
Based on linear systems theory, steady-state dynamics allow the nonlinear system \eqref{eq:sys} to be linearized around the equilibrium point \((\boldsymbol{x}_e, \boldsymbol{u}_e)\) through a first-order Taylor expansion \cite{cook2012flight}. Since by definition, quadrotor hovering balances the net forces and net moments into zero, the following approximation holds around the hovering equilibrium point
\begin{align}
\dot{\boldsymbol{x}}(t) \approx \mathbf{A}(\boldsymbol{x}_e, \boldsymbol{u}_e) (\boldsymbol{x}(t) - \boldsymbol{x}_e) + \mathbf{B}(\boldsymbol{x}_e, \boldsymbol{u}_e) (\boldsymbol{u}(t) - \boldsymbol{u}_e) , \label{eq:linearization}
\end{align}
where \( \boldsymbol{x}(t) \) is the state vector, \( \boldsymbol{u}(t) \) the control inputs, and the deviations from equilibrium are expressed as
\begin{align}
\delta \boldsymbol{x}(t) = \boldsymbol{x}(t) - \boldsymbol{x}_e \ \ \text{and} \ \ \delta \boldsymbol{u}(t) = \boldsymbol{u}(t) - \boldsymbol{u}_e \ .
\end{align}
Assuming linear time-invariant (LTI) dynamics in the vicinity of the equilibrium point, the system Jacobians with respect to the state and input are defined as
\begin{align}
\mathbf{A}(\boldsymbol{x}_e, \boldsymbol{u}_e) = \frac{\partial \boldsymbol{f}}{\partial \boldsymbol{x}} \bigg|_{(\boldsymbol{x}_e, \boldsymbol{u}_e)} 
\, \text{and} \ \,
\mathbf{B}(\boldsymbol{x}_e, \boldsymbol{u}_e) = \frac{\partial \boldsymbol{f}}{\partial \boldsymbol{u}} \bigg|_{(\boldsymbol{x}_e, \boldsymbol{u}_e)} , \label{eq:Jacobians}
\end{align}
reducing the error dynamics to the standard LTI form
\begin{equation} \label{eq:state_space}
\begin{aligned}
\dot{\boldsymbol{x}}(t) &= \mathbf{A} \, \boldsymbol{x}(t) + \mathbf{B} \, \boldsymbol{u}(t) \ , \\ 
\boldsymbol{y}(t) &= \mathbf{C} \, \boldsymbol{x} (t) \ ,
\end{aligned}
\end{equation}
where matrix $\mathbf{C}$ extracts the measurable states from $\boldsymbol{x}$. Model discrepancies and unmodeled dynamics are represented by zero-mean white Gaussian noise processes, with process covariance $\mathbf{W}$ and measurement covariance $\mathbf{V}$
\begin{align}
\boldsymbol{w}(t) &\sim \mathcal{N}(\mathbf{0}, \mathbf{W}) \ , \quad \mathbb{E}[\boldsymbol{ww}^{\top}] = \mathbf{W} \succ 0 \ , 
\\
\boldsymbol{v}(t) &\sim \mathcal{N}(\mathbf{0}, \mathbf{V}) \ \ , \quad \mathbb{E}[\boldsymbol{vv}^{\top}] = \mathbf{V} \succeq 0 \ . 
\end{align}
These propagate the open-loop uncertainty, whose state covariance ${\mathbf{P}}(t)$ evolves by the continuous-time Lyapunov equation
\begin{align}
\dot{\mathbf{P}}(t) = \mathbf{A} \mathbf{P}(t) + \mathbf{P}(t) \mathbf{A}^\top + \mathbf{W} . \label{eq:state_cov}
\end{align}
At any fixed time $t$, the joint distribution of the states and outputs can thus be approximated as Gaussian, namely
\begin{align}
\begin{bmatrix}
\boldsymbol{x}(t) \\
\boldsymbol{y}(t)
\end{bmatrix}
\hspace{-.5mm} \sim \mathcal{N} \hspace{-.5mm} 
\left(
\begin{bmatrix}
\boldsymbol{\mu}(t) \\
\mathbf{C} \boldsymbol{\mu}(t)
\end{bmatrix} 
, 
\begin{bmatrix}
\mathbf{P}(t) & \mathbf{P}(t) \mathbf{C}^\top \\
\mathbf{C} \mathbf{P}(t) & \mathbf{C} \mathbf{P}(t) \mathbf{C}^\top + \mathbf{V}
\end{bmatrix}
\right) . \label{eq:space}
\end{align}
where $\boldsymbol{\mu}(t)$ is the nominal noise-free state trajectory, initialized as $\boldsymbol{\mu} (t_0) = \mathbb{E}[\boldsymbol{x}(t_0)]$, with time evolution given by
\begin{align}
\dot{\boldsymbol{\mu}}(t) &= \mathbf{A} \boldsymbol{\mu}(t) + \mathbf{B} \boldsymbol{u}(t) \ . \label{eq:mu_dot}
\end{align}
%

\subsection{Inner-Level Optimization} 
Having identified the hovering operating point as locally linear under small perturbations enables the application of linear control theory. In this framework, solving the Riccati equations yields optimal controller ($\mathbf{K}$) and estimator ($\mathbf{L}$) gains, whereby the Kalman filter provides state estimates to the linear quadratic regulator (LQR). These gains are obtained by minimizing a quadratic cost over a finite time horizon $T$, 
\begin{align} \label{eq:cost}
\min_{\boldsymbol{u}(t)} \mathcal{J}_{\text{in}} \triangleq \lim_{T \rightarrow \infty} \mathbb{E} \left[ \frac{1}{T} \int_0^T ( \| \boldsymbol{x}(t) \|^2_{\mathbf{Q}} + \| \boldsymbol{u}(t) \|^2_{\mathbf{R}} ) \ \mathrm{d}t \, \right] ,
\end{align}
subject to \eqref{eq:mu_dot}, with a-priori specified \(\mathbf{Q} \succeq 0\) and \(\mathbf{R} \succ 0\) weighting deviations in $\boldsymbol{x}(t)$ and $\boldsymbol{u}(t)$, respectively \cite{aastrom2012introduction, chrif2014aircraft}. 
By defining the estimation and the control errors as
\begin{align}
\boldsymbol{e}(t) &= \boldsymbol{x}(t) - \hat{\boldsymbol{x}}(t) \ , \label{eq:err_est}
\\ 
\hat{\boldsymbol{\varepsilon}}(t) &= \hat{\boldsymbol{x}}(t) - \boldsymbol{x}_{\text{ref}} \ , \label{eq:err_ctr} 
\end{align}
the plant-estimator dynamics admits the block-state form of
\begin{align}
\begin{bmatrix}
    \dot{\boldsymbol{x}} \\ \dot{{\boldsymbol{e}}}
\end{bmatrix} = \begin{bmatrix}
    \mathbf{A-B K} & \mathbf{B K} \\ \boldsymbol{0} & \mathbf{A-LC}
\end{bmatrix} \begin{bmatrix} \boldsymbol{x} \\ {\boldsymbol{e}} \end{bmatrix} + \begin{bmatrix} \mathbf{I} & \boldsymbol{0} \\ \mathbf{I} & -\mathbf{L} \end{bmatrix} \begin{bmatrix} \boldsymbol{w} \\ \boldsymbol{v} \end{bmatrix} \ , \label{eq:error}
\end{align}
highlighting the separation principle, with an upper-triangular form that decouples the estimator dynamics \cite{khalil1996robust, mahmudov2000controllability, engelsman2025inertial}.
%

\subsection{System Dynamics}
We begin by outlining the assumptions and coordinate frames relevant to hovering, a near-equilibrium condition that justifies standard modeling practices \cite{hoffmann2007quadrotor, sadr2014dynamics}: (i) the quadrotor is a rigid symmetric body; (ii) thrust and moments scale with the square of rotor speeds; (iii) aerodynamic drag is neglected due to low speed; (iv) the center of mass coincides with the body-frame origin.
%
The orientation of the body frame \(\mathcal{B}\) relative to the inertial frame \(\mathcal{I}\) is given by the rotation matrix  
\begin{align}
\mathcal{R}_{\mathcal{B}}^{\mathcal{I}} := \left( \, \mathbf{e}_x^\mathcal{B}, \, \mathbf{e}_y^\mathcal{B}, \, \mathbf{e}_z^\mathcal{B} \, \right) \in SO(3) \ , \label{eq:frames}
\end{align}  
which is constructed through a sequence of three consecutive rotations: \(\mathcal{R}_z(\psi)\) about the \(\mathbf{e}_z^\mathcal{I}\) axis, \(\mathcal{R}_y(\theta)\) about the \(\mathbf{e}_y^\mathcal{I}\) axis, and \(\mathcal{R}_x(\phi)\) about the \(\mathbf{e}_x^\mathcal{I}\) axis, where \(\{\phi, \theta\} \in \left( -\frac{\pi}{2}, \frac{\pi}{2}\right)\) and \( \psi \in \left( -\pi, \pi \right)\) correspond to the roll, pitch, and yaw angles, respectively.
%
Building on the Newton-Euler formalism \cite{castillo2005modelling}, the coupled translational and rotational dynamics are expressed as
\begin{align} \label{eq:Newton_Euler}
\begin{pmatrix} 
\boldsymbol{f}^\mathcal{B} \\ \boldsymbol{\tau}^\mathcal{B} 
\end{pmatrix}=
\begin{pmatrix} 
m \boldsymbol{I}_{3} & \boldsymbol{0}_{3} \\ \boldsymbol{0}_{3} & \boldsymbol{J}
\end{pmatrix}
\begin{pmatrix}
\dot{\boldsymbol{v}}^\mathcal{B} \\  \dot{\boldsymbol{\omega}}^\mathcal{B}
\end{pmatrix}
+
\begin{pmatrix} 
\boldsymbol{\omega}^\mathcal{B} \times m \, \boldsymbol{v}^\mathcal{B} \\
\boldsymbol{\omega}^\mathcal{B} \times \boldsymbol{J} \boldsymbol{\omega}^\mathcal{B} 
\end{pmatrix} \, ,
\end{align}
where, $m$ is the body mass; $\boldsymbol{J}$, $\boldsymbol{I}_3$, and $\boldsymbol{0}_3$ are \( 3 \times 3 \) inertia tensor, identity, and zero matrices. Forces $\boldsymbol{f}$ and moments $\boldsymbol{\tau}$, expressed in the body frame, govern the time derivatives of $\boldsymbol{v}^\mathcal{B}$ and $\boldsymbol{\omega}^\mathcal{B}$.
%
In fixed-point hover, near-zero body velocities make drag negligible, so the net body-frame force includes only gravity $\mathbf{g}^\mathcal{B} = \text{g}\,\mathbf{e}_z^\mathcal{B}$ and vertical rotor thrusts $\boldsymbol{f}_{z,i}^\mathcal{B}$, yielding
\begin{align} \label{eq:thrust}
\boldsymbol{f}^\mathcal{B} = \sum_{i=1}^4 \boldsymbol{f}_{z,i}^\mathcal{B} - m \, \textbf{\textit{g}}^\mathcal{B} \ .
\end{align}
Each rotor spins along the body-frame $z$-axis, generating thrust by the $i$-th propeller modeled as
\begin{align} \label{eq:thrust_single}
\boldsymbol{f}^\mathcal{B}_{z,i} = {f}_{z,i} \, \textbf{e}^\mathcal{B}_z = \left( k_T \Omega_i^2 \right) \textbf{e}^\mathcal{B}_z \, ,
\end{align}
where $k_T$ is a lumped coefficient. Attitude stabilization is achieved through differential thrust between opposing rotors spaced by $\ell$. The rolling moment arises from the thrust difference between motors \#4 (left) and \#2 (right), given by
\begin{align}
\boldsymbol{\tau}_{\phi}^\mathcal{B} = \ell \left( \boldsymbol{f}^\mathcal{B}_{z,4} - \boldsymbol{f}^\mathcal{B}_{z,2} \right) \textbf{e}^\mathcal{B}_x \, ,
\end{align}
while the pitch moment is controlled by the longitudinal thrust difference between motors \#1 (front) and \#3 (rear), as
\begin{align}
\boldsymbol{\tau}_{\theta}^\mathcal{B} = \ell \left( \boldsymbol{f}^\mathcal{B}_{z,3} - \boldsymbol{f}^\mathcal{B}_{z,1} \right) \textbf{e}^\mathcal{B}_y \, .
\end{align}
%
Finally, yaw arises from asymmetry in rotor drag torques, i.e., when $\sum_i \tau_{M,i}\neq 0$. Defining $k_M$ as the torque coefficient and positive yaw as counter-clockwise about $\textbf{e}_z^{\mathcal{B}}$, the yaw torque is
\begin{align} \label{eq:drag_torque}
\boldsymbol{\tau}_{\psi}^\mathcal{B} = \sum_{i=1}^4 \tau_{M,i} = (-1)^i \left( k_M \Omega_i^2 \right) \textbf{e}^\mathcal{B}_z \, .
\end{align}
Fig.~\ref{fig:Quad_model} depicts the quadrotor free-body diagram: airframe (thick brown), reference axes (black), rotor thrust vectors (green), control inputs (azure), and rotor spin actuation (orange); gravity and inertial displacement are shown as dashed guides.
%
%
\tikzset{
  >={Stealth[length=5pt,width=5pt]},      
  every path/.style={line cap=round, line join=round},
  thickline/.style={line width=1.75pt},
  medline/.style={line width=1.5pt},
}

\colorlet{thrust}{orange}
\colorlet{spin}{green!90!cyan}
\colorlet{cmd}{blue!30!cyan}

\pgfmathsetmacro{\arcrx}{0.6}   
\pgfmathsetmacro{\arcry}{0.2}   
\pgfmathsetmacro{\uzscale}{1.4} 

\tikzset{
  arrow/.style={->, thickline},
  d_arrow/.style={->, medline, dashed},
  d_line/.style={medline, dashed},
  frame/.style={line width=4pt, draw=brown!80!black},
  axislabel/.style={font=\footnotesize, inner sep=1pt},
  circ/.style={circle, fill=black, minimum size=2.5pt, inner sep=0pt}
}


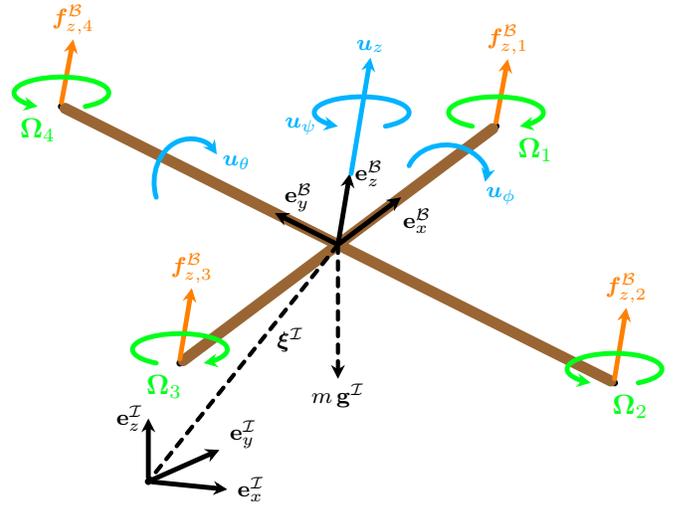
\begin{figure}[h]
\centering
\resizebox{\linewidth}{!}{%
\begin{tikzpicture}[scale=1.0, every node/.style={transform shape}]

\node[circ, shift={(+20mm,+15mm)}] at (0,0) (f_1) {};
\node[circ, shift={(+35mm,-17.5mm)}] at (0,0) (f_2) {};
\node[circ, shift={(-20mm,-15mm)}] at (0,0) (f_3) {};
\node[circ, shift={(-35mm,+17.5mm)}] at (0,0) (f_4) {};
\node[circ, shift={(-2.4,-3)}]       at (0,0) (cs_I) {};

\draw[frame] (f_1) -- (f_3);
\draw[frame] (f_2) -- (f_4);

\draw[arrow] (0,0) -- (0.8,0.6) node[axislabel,shift={(2mm,-3mm)}] {$\mathbf{e}_x^{\mathcal{B}}$};
\draw[arrow] (0,0) -- (-0.8,0.4) node[axislabel,shift={(3mm,2mm)}]  {$\mathbf{e}_y^{\mathcal{B}}$};
\draw[arrow] (0,0) -- (0.15,0.9) node[axislabel,right]               {$\mathbf{e}_z^{\mathcal{B}}$};

\draw[arrow,thrust] (f_1.north) -- ++(0.15,0.8) node[axislabel,above] {$\boldsymbol{f}^{\mathcal{B}}_{z,1}$};
\draw[arrow,spin]   ($(f_1)+(-0.3,0)$)
  arc[start angle=240,end angle=-60,x radius=\arcrx cm,y radius=\arcry cm]
  node[right=2mm,below] {$\boldsymbol{\Omega}_1$};

\draw[arrow,thrust] (f_2.north) -- ++(0.15,0.9) node[axislabel,above] {$\boldsymbol{f}^{\mathcal{B}}_{z,2}$};
\draw[arrow,spin]   ($(f_2)+(0.3,0)$)
  arc[start angle=-60,end angle=240,x radius=\arcrx cm,y radius=\arcry cm]
  node[right=5mm,below] {$\boldsymbol{\Omega}_2$};

\draw[arrow,thrust] (f_3.north) -- ++(0.15,0.9) node[axislabel,above] {$\boldsymbol{f}^{\mathcal{B}}_{z,3}$};
\draw[arrow,spin] ($(f_3)+(-0.3,0)$)
  arc[start angle=240,end angle=-60,x radius=\arcrx cm,y radius=\arcry cm]
  node[left=5mm,below] {$\boldsymbol{\Omega}_3$};

\draw[arrow,thrust] (f_4.north) -- ++(0.15,0.8) node[axislabel,above] {$\boldsymbol{f}^{\mathcal{B}}_{z,4}$};
\draw[arrow,spin]   ($(f_4)+(0.3,0)$)
  arc[start angle=-60,end angle=240,x radius=\arcrx cm,y radius=\arcry cm]
  node[left,below] {$\boldsymbol{\Omega}_4$};

\draw[arrow,cmd] (0.1666,0.9) -- ++(0.175*\uzscale,1.05*\uzscale)
  node[axislabel,above] {$\boldsymbol{u}_{z}$};

\draw[arrow,cmd] ($(0.6,1.5)$)
  arc[start angle=-60,end angle=240,x radius=\arcrx cm,y radius=\arcry cm]
  node[axislabel,above=1mm,left=2mm] {$\boldsymbol{u}_{\psi}$};

\draw[arrow,cmd] ($(-2.3,0.6)$)
  arc[start angle=200,end angle=40,x radius=0.45cm,y radius=0.55cm]
  node[axislabel,shift={(2.5mm,-1mm)}] {$\boldsymbol{u}_{\theta}$};

\draw[arrow,cmd] ($(0.9,1.)$)
  arc[start angle=160,end angle=-10,x radius=0.5cm,y radius=0.4cm]
  node[axislabel,right=2mm,below] {$\boldsymbol{u}_{\phi}$};

\draw[d_line]  (0,0) -- ++(-2.4,-3) node[axislabel,shift={(18mm,18mm)}] {$\boldsymbol{\xi}^{\mathcal{I}}$};
\draw[d_arrow] (0,0) -- (0,-17mm) node[axislabel,below] {$m\,\mathbf{g}^{\mathcal{I}}$};
\draw[arrow]   (cs_I) -- (-1.4,-3.1) node[axislabel,shift={(3mm,0mm)}]  {$\mathbf{e}_x^{\mathcal{I}}$};
\draw[arrow]   (cs_I) -- (-1.5,-2.6) node[axislabel,shift={(3mm,2mm)}]  {$\mathbf{e}_y^{\mathcal{I}}$};
\draw[arrow]   (cs_I) -- (-2.4,-2.2) node[axislabel,left]              {$\mathbf{e}_z^{\mathcal{I}}$};

\end{tikzpicture}%
}
\caption{Free-body diagram of the quadrotor system model.}
\label{fig:Quad_model}
\end{figure}

\subsection{System Kinematics}
The platform's trajectory is derived by time-integrating its dynamics and applying a coordinate transformation to an inertial reference frame, letting the position vector be \cite{valavanis2014handbook}
\begin{align}
{\boldsymbol{\xi}}^\mathcal{I} = \left(x, y, z \right)^\top \, ,
\end{align}
with the corresponding body velocities related by
\begin{align}
\dot{\boldsymbol{\xi}}^\mathcal{I} = \mathcal{R}_{\mathcal{B}}^{\mathcal{I}} {\boldsymbol{v}}^\mathcal{B} \, .
\end{align}
The Euler angles vector in the inertial frame is denoted as
\begin{align}
{\boldsymbol{\eta}}^\mathcal{I} = (\phi, \theta, \psi)^\top \, ,
\end{align}
with its rate of change related to the body angular rates \( (p, q, r) \) through the transformation matrix \( \textbf{\textit{W}}_{\boldsymbol{\eta}} \), such that
\begin{align}
\dot{\boldsymbol{\eta}}^\mathcal{I} = \textbf{\textit{W}}_{\boldsymbol{\eta}} \, {\boldsymbol{\omega}}^\mathcal{B} = 
\begin{pmatrix} 
1 & \text{s}_\phi \text{t}_\theta & \text{c}_\phi \text{t}_\theta \\
0 &  \text{c}_\phi & -\text{s}_\phi \\
0 & \text{s}_\phi / \text{c}_\theta & \text{c}_\phi / \text{c}_\theta 
\end{pmatrix} 
\begin{pmatrix} p \\ q \\ r \end{pmatrix} \, , \label{eq:W_eta}
\end{align}
where ’t’ is the tangent function. For derivation, thrust and torques are idealized as direct control input functions, hence
\begin{align} \label{eq:control}
\boldsymbol{u} (\Omega) = 
\begin{pmatrix}
\boldsymbol{u}_z \\ \boldsymbol{u}_\phi \\ \boldsymbol{u}_\theta \\ \boldsymbol{u}_\psi 
\end{pmatrix} 
= 
\begin{pmatrix}
{f}_{z} \, \textbf{e}_z^\mathcal{B} \\ 
{\tau}_{\phi} \, \textbf{e}_x^\mathcal{B} \\  
{\tau}_{\theta} \, \textbf{e}_y^\mathcal{B} \\  
{\tau}_{\psi} \, \textbf{e}_z^\mathcal{B}
\end{pmatrix} \in \ \mathbb{R}^4 \ .
\end{align}
Finally, by incorporating the generalized coordinates into the system model, the augmented state vector is defined as
\begin{align}
\boldsymbol{x} = \begin{pmatrix} \,
({\boldsymbol{\xi}}^\mathcal{I})^\top & ({\boldsymbol{v}}^\mathcal{B})^\top & ({\boldsymbol{\eta}}^\mathcal{I})^\top & ({\boldsymbol{\omega}}^\mathcal{B})^\top \
\end{pmatrix}^\top \, \in \mathbb{R}^{12} \, , \label{eq:state_vector}
\end{align}
with the nonlinear state-space model in \eqref{eq:sys} evolves as
\begin{align}
\begin{pmatrix}
\dot{\boldsymbol{\xi}}^\mathcal{I} \\ \dot{\boldsymbol{v}}^\mathcal{B} \\ \dot{\boldsymbol{\eta}}^\mathcal{I} \\ \dot{\boldsymbol{\omega}}^\mathcal{B} 
\end{pmatrix} = 
\begin{pmatrix}
(\mathcal{R}_{\mathcal{B}}^{\mathcal{I}}) \, {\boldsymbol{v}}^\mathcal{B} \\
m^{-1} \boldsymbol{u}_z - {\boldsymbol{\omega}}^\mathcal{B} \times {\boldsymbol{v}}^\mathcal{B} - {\textit{g}} \, \textbf{e}^\mathcal{B}_z \\ 
(\textbf{\textit{W}}_{\boldsymbol{\eta}}) \boldsymbol{\omega}^\mathcal{B} \\ 
\boldsymbol{J}^{-1} \left( ( \boldsymbol{u}_\phi + \boldsymbol{u}_\theta + \boldsymbol{u}_\psi ) - \boldsymbol{\omega}^\mathcal{B} \times \boldsymbol{J} \boldsymbol{\omega}^\mathcal{B} \right)
\end{pmatrix} . \label{eq:x_dot}
\end{align}

\section{Proposed Approach} \label{sec:prop}
We begin by analyzing the linearized dynamics about the hover equilibrium, outlining the outer evolutionary LQG optimization process, and lastly presenting our proposed implementation amongst several state-of-the-art tuning methods.

\subsection{Hovering Linearization}
Since hovering occurs near equilibrium with small state variables, local behavior can be approximated using small-angle assumptions. This simplifies \eqref{eq:W_eta} to $\dot{\boldsymbol{\eta}}^\mathcal{I} \approx \boldsymbol{\omega}^\mathcal{B}$ and $\mathcal{R}_{\mathcal{B}}^{\mathcal{I}} \approx \boldsymbol{I}_3 + (\boldsymbol{\eta})_{\times}$, respectively, where $(\cdot)_{\times}$ denotes the skew-symmetric operator. With these approximations, the Jacobian of the nonlinear dynamics in \eqref{eq:x_dot} is computed as
\begin{align}
\frac{\partial \boldsymbol{f}}{\partial \boldsymbol{x}} = 
\begin{pmatrix}
\boldsymbol{0}_3 & \boldsymbol{I}_{3} + (\boldsymbol{\eta})_\times & -({\boldsymbol{v}}^\mathcal{B})_\times & \boldsymbol{0}_3 \\ 
\boldsymbol{0}_3 & -({\boldsymbol{\omega}}^\mathcal{B})_\times & - (\textbf{\textit{g}}^\mathcal{I})_\times & ({\boldsymbol{v}}^\mathcal{B})_\times \\ 
\boldsymbol{0}_3 & \boldsymbol{0}_3 & \boldsymbol{0}_3 & \boldsymbol{I}_3 \\ 
\boldsymbol{0}_3 & \boldsymbol{0}_3 & \boldsymbol{0}_3 & \nabla_{\boldsymbol{\omega}} (\dot{\boldsymbol{\omega}}^\mathcal{B} )
\end{pmatrix} ,
\end{align}
with the corresponding sub-Jacobian of $\dot{\boldsymbol{\omega}}^\mathcal{B}$ given by
\begin{align}
\nabla_{\boldsymbol{\omega}} (\dot{\boldsymbol{\omega}}^\mathcal{B} ) = \boldsymbol{J}^{-1} \left( \, ( \boldsymbol{J} \boldsymbol{\omega}^\mathcal{B} )_{\times} - \boldsymbol{\omega}^\mathcal{B} \times \boldsymbol{J} \, \right) \, .
\end{align}
Assuming a hovering equilibrium defined by a state vector at a fixed point and a nominal thrust control input
\begin{align}
\boldsymbol{x}_e &= \begin{pmatrix} \boldsymbol{\xi}^\mathcal{I}_e & \mathbf{0}_3 & {\boldsymbol{\eta}}^\mathcal{I}_e & \mathbf{0}_3 \end{pmatrix}^\top \ \in \ \mathbb{R}^{12} \ , \label{eq:equilibrium_state} \\
\boldsymbol{u}_e &= \begin{pmatrix} m \textit{g} & 0 & 0 & 0 \end{pmatrix}^\top \ \in \ \mathbb{R}^{4} \ , \label{eq:equilibrium_input}
\end{align}
substituting this operating point into the state Jacobian yields the linearized system dynamics model
\begin{align} \label{eq:A_mat}
\mathbf{A} = \frac{\partial \boldsymbol{f}}{\partial \boldsymbol{x}}  \bigg|_{(\boldsymbol{x}_e, \boldsymbol{u}_e)} = \hspace{5cm} \\ \notag
\begin{pmatrix}
\boldsymbol{0}_3 & \boldsymbol{I}_3 + (\boldsymbol{\eta}_e)_\times & \boldsymbol{0}_3 & \boldsymbol{0}_3 \\ 
\boldsymbol{0}_3 & \boldsymbol{0}_3 & - (\textbf{\textit{g}}^\mathcal{I})_\times & \boldsymbol{0}_3 \\ 
\boldsymbol{0}_3 & \boldsymbol{0}_3 & \boldsymbol{0}_3 & \boldsymbol{I}_3 \\ 
\boldsymbol{0}_3 & \boldsymbol{0}_3 & \boldsymbol{0}_3 & \boldsymbol{0}_3 
\end{pmatrix} \in \, \mathbb{R}^{12 \times 12} ,
\end{align}
where higher-order centrifugal terms and their Coriolis couplings vanish. Likewise, linearizing the control input around the thrust required to balance gravity gives
\begin{table*}[b]
\centering
\caption{Comparative analysis of weight optimization methods, ordered top-to-bottom by their novelty and precision.}
\renewcommand{\arraystretch}{1.375}
\begin{tabular}{|c|l|l|l|l|c|}
\hline
\textbf{Symbol} \CC & \textbf{Optimization method} \CC & \textbf{Approach} \CC& \textbf{Pros} \CC & \textbf{Cons} \CC & \textbf{Complexity} \CC \\
\hline 
MT  & Manual tuning & Trial-and-error & Trivial implementation & Inefficient (exhaustive)   & ${O}(n\,f)$ \\ 
BR  & Bryson’s rule \cite{hespanha2018linear}& Heuristic & Intuitive; Fast initial guess      & Inherently suboptimal    & ${O}(f)$ \\
PS  & Particle swarm \cite{fessi2019lqg}& Swarm-based& Robust against local minima & Premature convergence & ${O}(k\,n\,f)$ \\
GA  & Genetic algorithm \cite{abdulla2023aircraft}& Evolutionary& Global search; gradient-free    & Computationally intensive   & ${O}(k\,n\,f)$  \\
BY & Bayesian \cite{marco2016automatic}& Surrogate-based& Sample-efficient& Scalability issues & ${O}(k\,f + k^3)$ \\ \hline
CMA \CG& Covariance matrix adaptation \cite{hansen2016cma} \CG& Simulation-based  \CG& Robust in rugged search spaces  \CG & Slower convergence \CG & ${O}(k\,n\,f)$ \CG \\
\hline 
\end{tabular}
\label{tab:lqg_tuning_methods}
\end{table*}
\begin{align}
\mathbf{B} = \frac{\partial \boldsymbol{f}}{\partial \boldsymbol{u}} \bigg|_{(\boldsymbol{x}_e, \boldsymbol{u}_e)} = \begin{pmatrix} 
\boldsymbol{0}_{3 \times 1} & \boldsymbol{0}_{3 \times 3} \\ 
m^{-1} \textbf{e}_z^{\mathcal{B}} & \boldsymbol{0}_{3 \times 3} \\ 
\boldsymbol{0}_{3 \times 1} & \boldsymbol{0}_{3 \times 3} \\ 
\boldsymbol{0}_{3 \times 1} & \boldsymbol{J}^{-1}
\end{pmatrix} \in \, \mathbb{R}^{12 \times 4} \, .
\end{align}
Within the state estimation process, these linearized dynamics hold locally around the hover equilibrium $(\boldsymbol{x}_e, \boldsymbol{u}_e)$, enabling state propagation via the state-space model \eqref{eq:state_space}.

\subsection{Outer-Level Optimization}
Gradient-based methods, while effective in high-dimensional settings, are unsuitable here since analytical gradients are unavailable: the cost depends implicitly on closed-loop dynamics. The LQG cost \eqref{eq:cost} is set a priori by $\mathbf{Q}$ and $\mathbf{R}$, whose iterative tuning forms an inner feedback loop within the outer optimizer. As hovering prioritizes position holding, followed by attitude and yaw stabilization, the outer cost is defined as
\begin{align}
\mathcal{J}_{\text{out}} \;=\; \int_0^T \Big( \,\lVert \boldsymbol{\varepsilon}_{\boldsymbol{\xi}} (t)\rVert_2 \;+\; \lambda \,\lVert \boldsymbol{\varepsilon}_{\boldsymbol{\eta}}(t)\rVert_2 \Big)\, \mathrm{d} t \, ,
\end{align}
where $\lambda \ll 1 $ penalizes the translational–rotational error trade-off. For an $n$-dimensional state-space model \eqref{eq:state_space}, the search spans $2n^2$ weight parameters, placing the bi-level tuning firmly in the domain of black-box optimization, where weights are iteratively refined without explicit gradients or access to internal system structure.
\\
\textit{Remark 1:} To enforce positive definiteness, both weight matrices are parameterized via Cholesky factorization, $\mathbf{Q} = \boldsymbol{L}_\text{Q} \boldsymbol{L}_\text{Q}^\top$, and $\mathbf{R} = \boldsymbol{L}_\text{R} \boldsymbol{L}_\text{R}^\top$, with $\boldsymbol{L}$ being lower triangular, effectively reducing the parameter count to $n(n+1)$.
\\
Algorithm \ref{alg:bb_opt} formalizes the proposed framework in a high-level, optimizer-agnostic form, with the inner LQG control loop nested within the outer weight-optimization process.
\begin{algorithm}[h]
\caption{AERO-LQG bi-level optimization structure}
\label{alg:bb_opt}
\KwIn{Optimizer $\mathcal{O}$, weight $\lambda$, tolerance $\epsilon$}
Initialize $\mathcal{O}$ and weights $(\mathbf{Q}, \mathbf{R})$\tcp*[r]{outer-loop}
\While{not converged}
{
    Propose candidate sets $\{(\mathbf{Q}_i, \mathbf{R}_i)\}$ using $\mathcal{O}$\;
    \ForEach{candidate $(\mathbf{Q}_i, \mathbf{R}_i)$}
    {
        $ \left( \boldsymbol{\xi}(t), \boldsymbol{\eta}(t) \right) \gets \arg\min \mathcal{J}_{\text{in}}$\tcp*[r]{inner-loop}
    }
    Feed optimal errors $(\boldsymbol{\varepsilon}_{\boldsymbol{\xi}}^*, \boldsymbol{\varepsilon}_{\boldsymbol{\eta}}^*)$ into optimizer $\mathcal{O}$\; 
    \If{$\mathcal{J}_{\text{out}} < \epsilon$}{
    \textbf{break}\tcp*[r]{convergence}
    }
    }
\Return $(\mathbf{Q}^*, \mathbf{R}^*)$\;
\end{algorithm}
%
\\
Table~\ref{tab:lqg_tuning_methods} lists the six optimization techniques used as benchmarks in this study, distinguished by their key characteristics and ranked in ascending order of complexity. To assess the effectiveness of our covariance matrix adaptation (CMA) implementation—highlighted in green at the bottom—complexity is expressed in terms of number of parameters ($n$), iterations ($k$), and cost per function evaluation ($f$).
\newpage

\section{Analysis and Results} \label{sec:results}
To explore the intricacies of LQG tuning in quadrotor hovering, AERO-LQG results are organized into three subsections.

\subsection{Cost landscape}
\begin{figure}[h] 
\begin{center}
\includegraphics[width=0.465\textwidth]{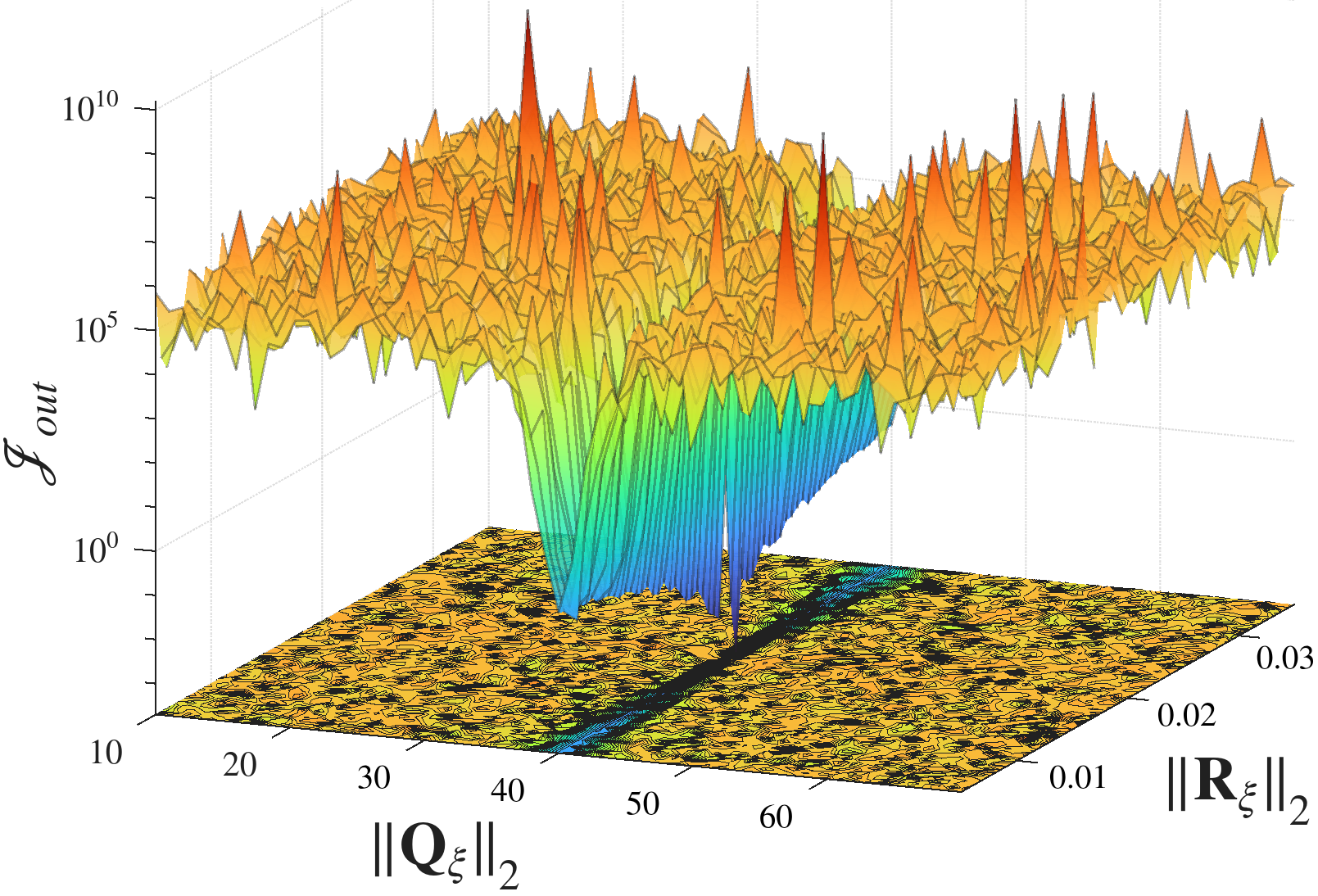}
\caption{LQG's marginal stability spawns a rugged, minima-rich cost surface, calling for gradient-free optimization.}
\label{fig:cost}
\end{center}
\end{figure}

When applied to a 12-dimensional state space rich in cross-coupled dynamics, the LQG quadratic cost \eqref{eq:cost} yields a highly non-convex landscape over the gain parameters. In such settings, gradient-based optimizers are prone to becoming trapped in inferior local minima, with little prospect of escape. To demonstrate, we compute the cost of a 10-second standardized gust trajectory, assuming isotropic 3D error and using L2 norms for $\mathbf{Q}_{\xi}$ and $\mathbf{R}_{\xi}$ for clarity. As shown in Fig.~\ref{fig:cost}, the scalar surface is riddled with local minima, exposing the shortcomings of gradient-based techniques.
\\
This, in turn, accentuates the challenge of proper LQG tuning: large regions of the cost surface correspond to poor controller performance, with log-scale costs on the order of $\mathcal{J}_{\text{out}} \gtrsim 10^5$. Yet, precise calibration of $\mathbf{Q}_{\xi}$ and $\mathbf{R}_{\xi}$ norms reduces the cost by nearly seven orders of magnitude, achieving near-global optimality and substantially improved performance.
\addtocounter{figure}{+1}
\begin{figure*}[b]
\centering
\includegraphics[width=.99\textwidth]{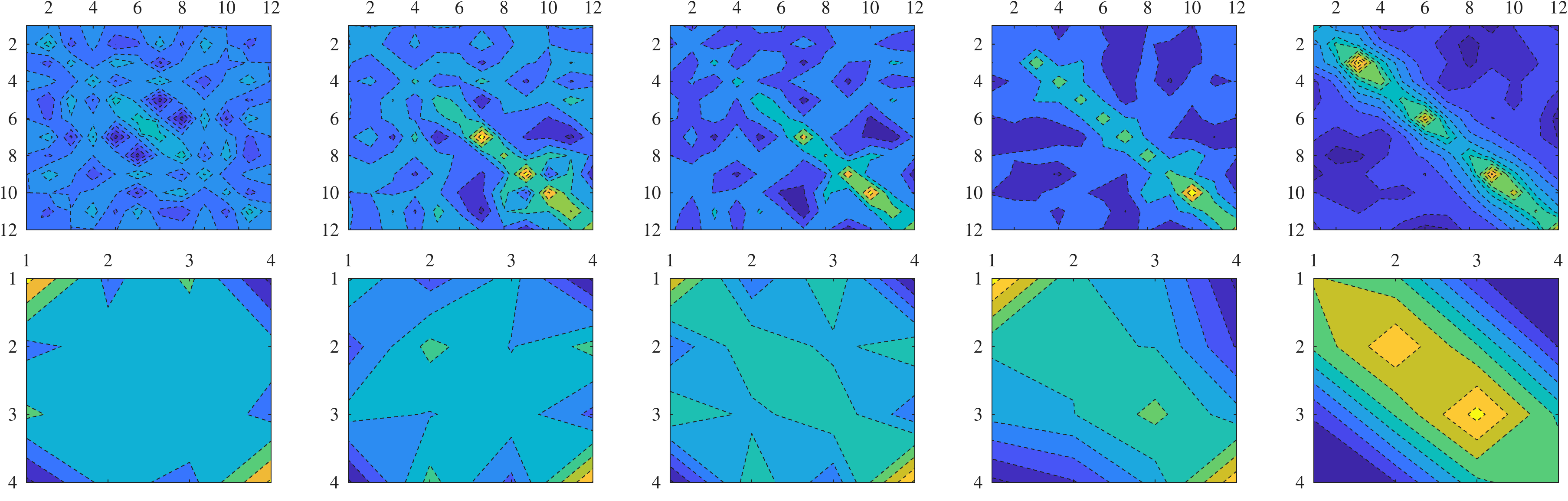}
\caption{Evolutionary optimization guides $\mathbf{Q}$ (top) and $\mathbf{R}$ (bottom) from random start (left) toward diagonal dominance (right).}
\label{fig:covariance}
\end{figure*}

\subsection{LQG inherent instability}
Fig.~\ref{fig:res_unstable} illustrates the onset of instability in quadrotor states over a 10-second standardized gust trajectory \cite{engelsman2025c}, with the 3D error norms from \eqref{eq:err_est} and \eqref{eq:err_ctr} on the left, and all four control inputs \eqref{eq:control} on the right. Even under careful manual tuning, the system diverges. In delicate hovering, LQG’s narrow stability margins make it highly sensitive to small disturbances or accumulated noise; once error assumptions are violated, divergence grows unchecked, and the actuators respond with rapid but ineffective corrections, oscillating between their mechanical limits.
\addtocounter{figure}{-2}
\begin{figure}[h]
\begin{center}
\includegraphics[width=0.475\textwidth]{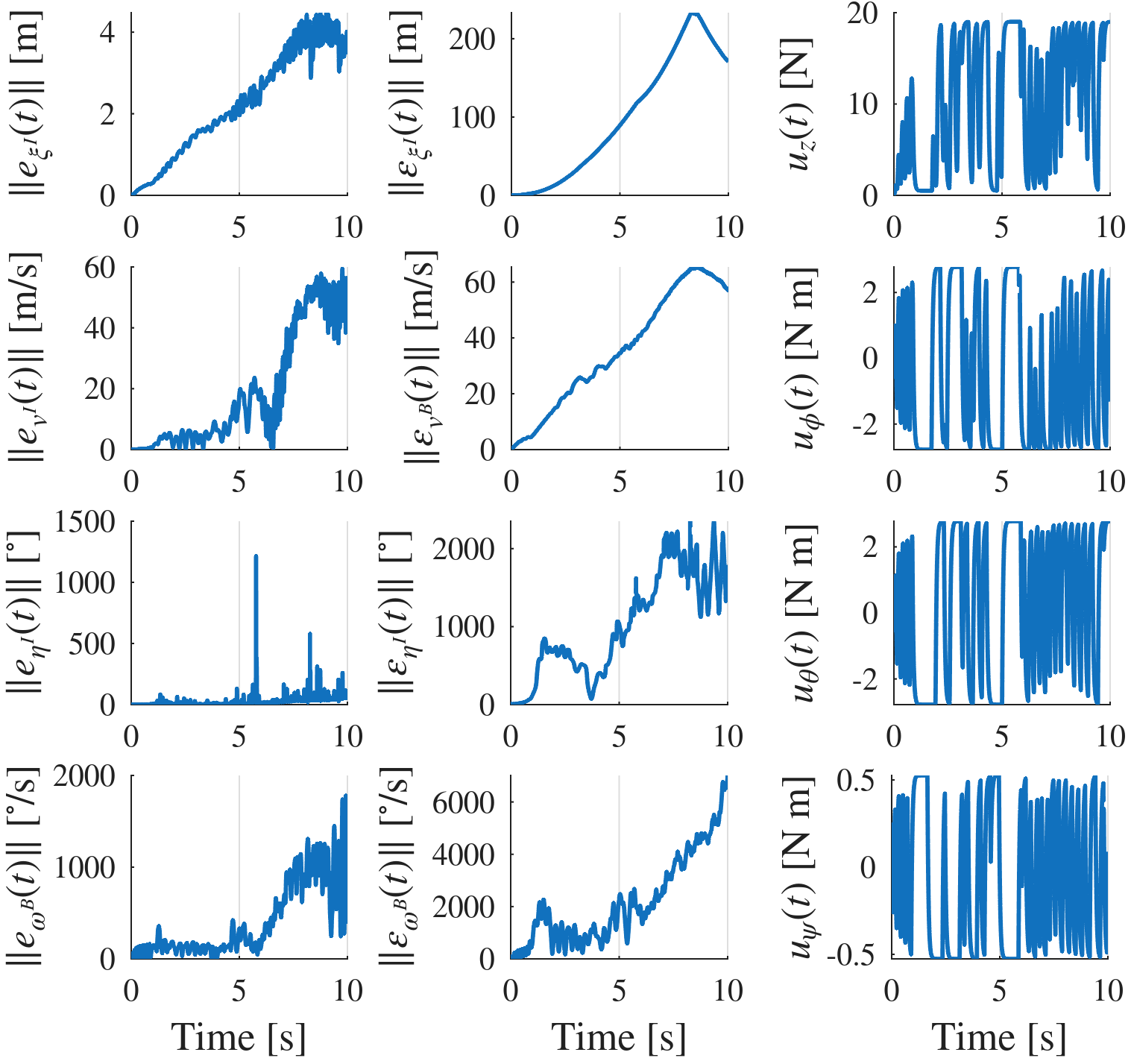}
\caption{Quadrotor state trajectories under LQG instability.}
\label{fig:res_unstable}
\end{center}
\end{figure}
\\
This instability arises from the tight coupling between the estimator and controller, where even small deviations can degrade the Kalman filter, initiating a feedback loop that amplifies errors. Unlike LQR, which acts directly on measured states, LQG relies on state estimates, making it inherently more sensitive to estimation inaccuracies and tuning imperfections.

\subsection{Performance Evaluation}
Building on the need for robust, global optimization methods that withstand large perturbations, this section: (i) presents our CMA implementation, (ii) analyzes its performance, and (iii) benchmarks it against several state-of-the-art approaches.
\\
Fig.~\ref{fig:covariance} illustrates CMA’s iterative refinement across five stages, guiding the search distribution from random initialization (left) to convergence with the predefined criteria (right).
\addtocounter{figure}{+1}
\begin{figure}[h]
\begin{center}
\includegraphics[width=0.5\textwidth]{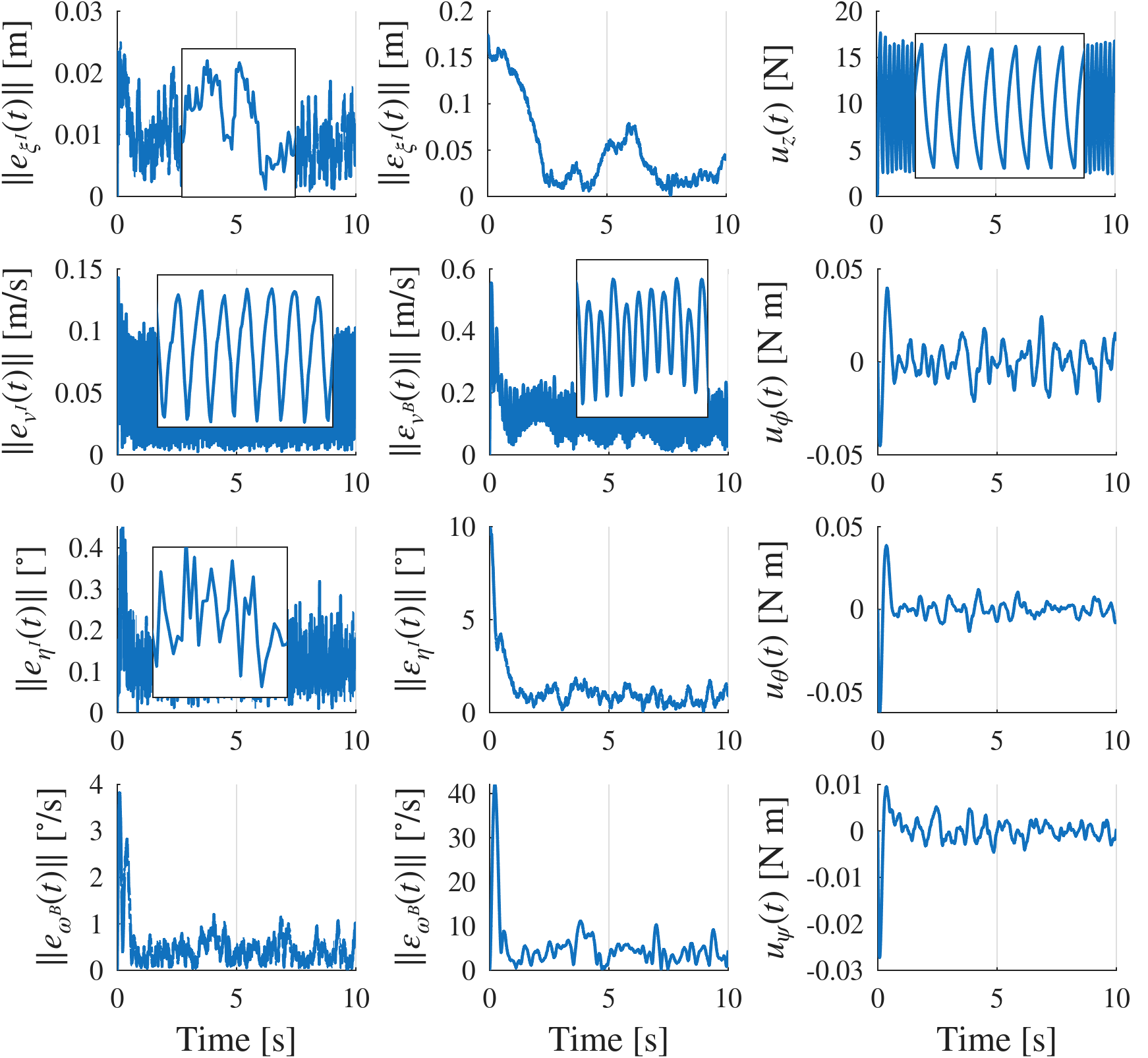}
\caption{Quadrotor state trajectories under LQG stability.}
\label{fig:res_stable}
\end{center}
\end{figure}
\\
Normalized for convenience, both $\mathbf{Q}$ and $\mathbf{R}$ progressively concentrate their weight along the diagonal, while off-diagonal entries diminish—indicating negligible cross-correlation.
\begin{table}[b]
\centering
\caption{Comparison of optimizers by final-time errors.}
\renewcommand{\arraystretch}{1.6}
\begin{tabular}{|c|c|c|c|c|c|}
\hline 
\multirow{2}{*}{\diagbox[width=19mm]{Model}{Error}} & \multicolumn{2}{c|}{Position [m] \CC} & \multicolumn{2}{c|}{Orientation [deg] \CC} & Effort [N$\cdot$s] \CC \\ \cline{2-6}
& $\| \boldsymbol{e}_{\boldsymbol{ \xi}^{\mathcal{I}}} \|$& $\| \boldsymbol{\varepsilon}_{\boldsymbol{ \xi}^{\mathcal{I}}} \|$ & $\| \boldsymbol{e}_{\boldsymbol{ \eta}^{\mathcal{I}}} \|$ & $\| \boldsymbol{\varepsilon}_{\boldsymbol{ \eta}^{\mathcal{I}}} \|$ & $\int_0^T \boldsymbol{u}(t) \, \mathrm{d}t$ \\ \hline \hline 
MT\CC & $\to \infty$ & $\to \infty$ & $\to \infty$ & $\to \infty$ & $\to \infty$ \\ \hline 
BR\CC & 0.48 & 1.12 & 0.57 & 12.3 & $> 10^3$ \\ \hline 
PS\CC & 0.22 & 0.77 & 0.18 & 6.13 & 293.8 \\ \hline 
GA\CC & 0.03 & 0.09 & 0.22 & 2.05 & 119.1 \\ \hline 
BY\CC & 0.31 & 0.27 & 0.47 & 3.21 & 143.1 \\ \hline 
CMA\CG & \textbf{0.01} \CG& \textbf{0.05} \CG & \textbf{0.17} \CG& \textbf{1.74} \CG& \textbf{98.6} \CG \\ \hline 
\end{tabular} \label{t:comp_stability}
\end{table}
\\
Fig.~\ref{fig:res_stable} follows the same top-to-bottom error analysis—position, velocity, orientation, and angular velocity—under the finalized parameters. In contrast to Fig.~\ref{fig:res_unstable}, where saturated controllers failed to restore stability, the trajectories here recover immediately from the same unit disturbance. A residual drift persists, visible as sawtooth patterns in the insets; however, the actuators remain within limits, achieving stabilization with significantly lower control effort (rightmost column).
\\
Table~\ref{t:comp_stability} provides supporting numbers: MT at the top exhibits complete divergence ($\infty$), illustrating why manual tuning falls short next to automated algorithms. In contrast, the optimizers below perform far better, with CMA surpassing the leading GA by reducing estimation and control errors by at least 33\% and 55\%, respectively. Fig.~\ref{fig:benchmarking} shows these differences across eight flight metrics from \cite{engelsman2025c}, benchmarking all candidate optimizers (excluding MT) with color-coded distinction.
\begin{figure}[h]
\begin{center}
\includegraphics[width=0.487\textwidth]{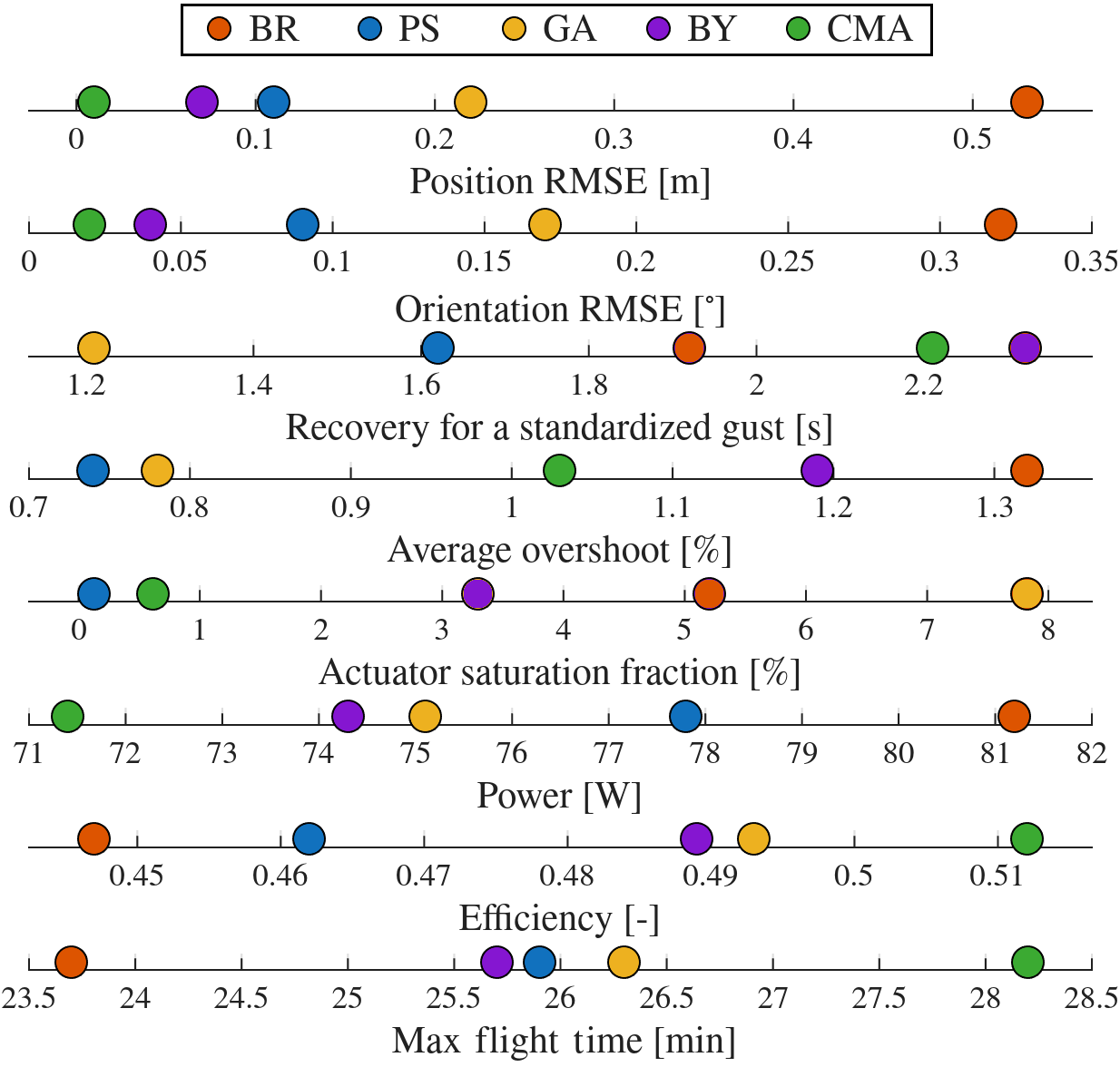}
\caption{Head-to-head comparison of optimizers' performances.}
\label{fig:benchmarking}
\end{center}
\end{figure}
\\
As shown, CMA (green) attains optimality in five of eight criteria, excelling in precision and energy efficiency but lagging in agility and responsiveness. Overall, evolutionary strategies (PS, GA, CMA) consistently outperform heuristic methods (MT, BR) and Bayesian optimization by avoiding premature convergence in dense local minima. These advantages provide tangible benefits, particularly in resource-constrained applications where extended flight endurance is critical.
%

\section{Discussion and Conclusion} \label{sec:conc}
This work introduced AERO-LQG, a systematic weight-tuning framework for quadrotor hovering that addresses the narrow stability margins and highly non-convex cost landscapes characteristic of such systems. By decoupling the inner control loop from the outer stochastic optimizer, weight initialization is delegated to a powerful evolutionary strategy that remains robust to complex cost landscapes. Benchmarking across eight flight metrics showed the consistent superiority of our CMA implementation in key hovering capabilities—energy efficiency, flight endurance, and disturbance rejection—a result particularly significant for resource-constrained platforms. While demonstrated on an aerial system, the approach generalizes to other robots, offering resource-aware, high-performance control for real-world demands.

\bibliographystyle{IEEEtran}
\bibliography{Ref}
\end{document}